# Tunable Doping of Rhenium and Vanadium into Transition Metal Dichalcogenides for Two-Dimensional Electronics

*Shisheng Li,[*] Jinhua Hong, Bo Gao, Yung-Chang Lin, Hong En Lim, Xueyi Lu, Jing Wu, Song Liu, Yoshitaka Tateyama, Yoshiki Sakuma, Kazuhito Tsukagoshi, Kazu Suenaga and Takaaki Taniguchi[*]*

Dr. S. Li
International Center for Young Scientists (ICYS), National Institute for Materials Science (NIMS), Tsukuba 305-0044, Japan
E-mail: LI.Shisheng@nims.go.jp

Dr. J. Hong, Dr. Y.-C. Lin, Prof. K. Suenaga
Nanomaterials Research Institute, National Institute of Advanced Industrial Science and Technology, AIST Central 5, Tsukuba 305-8564, Japan

Dr. B. Gao, Dr. Y. Tateyama
Center for Green Research on Energy and Environmental Materials (GREEN), National Institute for Materials Science (NIMS), Tsukuba 305-0044, Japan

Dr. H. E. Lim
Department of Physics, Tokyo Metropolitan University, Hachioji 192-0397, Japan

Dr. J. Wu
Institute of Materials Research and Engineering, Agency for Science, Technology and Research, 138634, Singapore

Prof. S. Liu
Institute of Chemical Biology and Nanomedicine (ICBN), College of Chemistry and Chemical Engineering, Hunan University, Changsha 410082, P. R. China

Dr. B. Gao, Dr. X. Lu, Dr. Y. Tateyama, Dr. K. Tsukagoshi, Dr. T. Taniguchi
International Center for Materials Nanoarchitectonics (WPI-MANA), National Institute for Materials Science (NIMS), Tsukuba 305-0044, Japan
E-mail: TANIGUCHI.Takaaki@nims.go.jp

Dr. Y. Sakuma
Research Center for Functional Materials, National Institute for Materials Science (NIMS), Tsukuba 305-0044, Japan

Abstract

Two-dimensional (2D) transition metal dichalcogenides (TMDCs) with unique electrical properties are fascinating materials used for future electronics. However, the strong Fermi level pinning effect at the interface of TMDCs and metal electrodes always leads to high contact resistance, which seriously hinders their application in 2D electronics. One effective way to overcome this is to use metallic TMDCs or transferred metal electrodes as van der Waals (vdW) contacts. Alternatively, using highly conductive doped TMDCs will have a profound impact on the contact engineering of 2D electronics. Here, a novel chemical vapor deposition using mixed molten salts is established for vapor-liquid-solid growth of high-quality rhenium (Re) and vanadium (V)-doped TMDC monolayers with high controllability and reproducibility. A tunable semiconductor to metal transition is observed in the Re and V-doped TMDCs. Electrical conductivity increases up to a factor of $10^8$ in the degenerate V-doped $WS_2$ and $WSe_2$. Using V-doped $WSe_2$ as vdW contact, the on-state current and on/off ratio of $WSe_2$-based field effect transistors have been substantially improved (from ~$10^{-8}$ to $10^{-5}$ A; ~$10^4$ to $10^8$), compared to metal contacts. Future studies on lateral contacts and interconnects using doped TMDCs will pave the way for 2D integrated circuits and flexible electronics.

**Main Text**

Two-dimensional (2D) materials, e.g., graphene, h-BN and transition metal dichalcogenides (TMDCs), are important building blocks for future electronics and optoelectronics. Among them, the group VI semiconducting TMDCs are potential candidates for the channel materials of field effect transistors (FETs) as they exhibit reasonable carrier mobility (tens to hundreds $cm^2/Vs$) and high current on/off ratio (up to $10^8$).[1-4] Furthermore, they are endowed with mechanical flexibility and high optical transparency for applications in next-generation, light-weight, flexible and wearable electronics.[5] However, one serious issue regarding 2D TMDC-based electronics is to overcome the strong Fermi level pinning effect that fixes Schottky barrier heights (SBHs) at the TMDC/metal contact interface.[6-12] It hinders carrier modulation, causing high contact resistance in the TMDC-based devices as a consequence. To solve this problem, numerous efforts have been devoted to the contact engineering. For instance, vertical or lateral contact has been established with metallic 2D materials.[13-24] Besides, instead of direct metal deposition, prepatterned metal electrodes are transferred as contacts to avoid damages or strain to the 2D TMDCs, which cause high contact resistance.[25] Meanwhile, ion implantation is a well-established doping technique to modulate the carrier density and to realize ohmic contact in silicon-based devices. Similar, substitutional doping of 2D TMDCs provides a brand-new strategy for the contact engineering and carrier modulation of 2D TMDC-based electronics.[26, 27]

In the last five years, many important advances in TMDC doping have been achieved by chemical vapor deposition (CVD) method. These include direct growth with a mixture of transition metal oxides/halides,[26, 28-31] liquid-phase precursors[32-34] or metal organic reactants (MOCVD),[35] which remarkably enhance the electrical, magnetic and catalytic properties of doped TMDCs. However, vaporization of the transition metal precursors from solid powders oftentimes leads to poor uniformity and controllability of as-grown doped TMDCs. Furthermore, the high concentration of transition metal precursors and byproducts has also caused severe contaminations. Although the MOCVD shows good uniformity and controllability, as-grown doped TMDCs still suffers from small grain size (sub-micrometer level), not to mention the high cost involved. Therefore, a new synthetic technique compatible with conventional ambient-pressure CVD, which can be handled with great feasibility and easy controllability, is highly demanding to precisely tune the doping and electrical properties of as-grown TMDCs. Only thus can doped TMDCs make substantial contribution to future 2D electronics.

In this communication, we report a versatile, highly reproducible vapor-liquid-solid (VLS) growth of Re and V-doped TMDC monolayers using the molten salts, $Na_2MoO_4$, $Na_2WO_4$,

NaReO$_4$ and NaVO$_3$. By adjusting the NaReO$_4$ and NaVO$_3$ ratio in their mixtures with Na$_2$MoO$_4$ or Na$_2$WO$_4$, high-quality Re and V-doped TMDC (MoS$_2$, WS$_2$ and WSe$_2$) monolayers with tunable composition can be easily grown by sulfurization and selenization. With the use of molten salts, the mixed salts are in liquid state during the TMDC growth. The mixture is thus has a higher spatial and composition uniformity compared to the vaporized transiton metal precursors. Besides, it has minimal contamination to the reaction chamber. Repeated growth of more than a hundred batches can be conducted using a single growth chamber under ambient pressure CVD process. We attribute this to the thin-layer, spin-coated molten salts used for growing Re, and V-doped TMDCs. This is almost impossible for the doped TMDCs grown with vaporized precursors. In addition, the as-grown Re and V-doped TMDCs can easily achieve a very large domian size up to 100 μm. Our spectroscopic studies using Raman and photoluminescence (PL) along with the scanning transmission electron microscopy (STEM) observations confirm the successful doping of Re and V in the as-grown samples. The electrical properties of these Re and V-doped TMDC-based FETs show possible carrier modulation. Distinct semiconductor to metal transition is observed with the increase of Re and V concentration. Highly conductive V-doped WSe$_2$ were employed as vertical vdW contact for WSe$_2$-FETs. Compared to typical metal contacts using Au and Pd, the on-state current and on/off ratio of WSe$_2$-based FETs have been substantially improved from ~$10^{-6}$-$10^{-8}$ A to $10^{-5}$ A and from ~$10^4$ to $10^8$, respectively. Our results demonstrate the great potentials of Re and V-doped TMDCs for future 2D TMDC-based electronics.

(**Figure 1**)

To modify the electrical properties of group VI TMDCs, one effective way is via substitutional replacement of Mo and W elements with neighboring Re (electron donor), V or Nb (electron acceptor) in periodic table (**Figure 1**a). The band structure, carrier type and electrical properties of 2D TMDCs can be systematically tuned with the dopant concentration.

Our recent achievements in molten-salt CVD, VLS growth of TMDCs show great promise for synthesizing doped TMDCs with high controllability and great feasibility.[36, 37] Molten slats, e.g., $Na_2MoO_4$ and $Na_2WO_4$ have melting points of 687 °C and 698 °C, respectively. They are in liquid (molten) state because of their low vapor pressures at the typical growth temperatures of TMDC monolayers, around 750 °C. In stark contrast to the volatile transition metal oxides, chlorides and oxychlorides, which grow TMDC monolayers in a vapor-solid-solid (VSS) mechanism,[4, 38] these molten salts grow TMDC monolayers in a VLS mechanism.[36, 37] In principle, mixed molten salts are expected to have better controllability than the vapor precursors in growing doped TMDCs. Here, we have chosen similar molten salts, $NaReO_4$ ($T_m$ = 414 °C) and $NaVO_3$ ($T_m$ = 630 °C) to grow the Re and V-doped TMDC monolayers, respectively. First, 20 mM $Na_2MoO_4$, $Na_2WO_4$, $NaReO_4$ and $NaVO_3$ aqueous solution were prepared as source precursors. Then, the mixed salt solutions with different $NaReO_4$ and $NaVO_3$ ratios were prepared as illustrated in Figure 1b. Here and thereafter, $X$ denotes the mixed salts of $Na_2MoO_4$-$NaReO_4$ ($X_{ReMo}$), $Na_2MoO_4$-$NaVO_3$ ($X_{VMo}$), $Na_2WO_4$-$NaReO_4$ ($X_{ReW}$) and $Na_2WO_4$-$NaVO_3$ ($X_{VW}$). $X^{Re}$ and $X^V$ represent the ratio of $NaReO_4$ and $NaVO_3$ in mixed salts, respectively. Later, the mixed salt solutions were spin-coated onto sapphire substrates. Finally, Re and V-doped TMDC monolayers were grown by either sulfurizing or selenizing the mixed salts in an ambient-pressure thermal CVD system (Figure 1c).

(**Figure 2**)

To investigate the growth of tunable Re and V-doped TMDC monolayers, the $Na_2MoO_4$-$NaReO_4$ ($X_{ReMo}$) and $Na_2MoO_4$-$NaVO_3$ ($X_{VMo}$) mixed salts were first utilized for the CVD growth of Re and V-doped $MoS_2$ monolayers, respectively. Our X-ray photoelectron spectroscopy (XPS) results demonstrate successful Re doping in the as-grown $MoS_2$ monolayers (Figure S1). The optical images of the as-grown Re and V-doped $MoS_2$ monolayers are shown in Figure S2. Overall, the grain size and surface coverage of the as-grown Re and V-doped MoS$_2$ monolayers show an obvious decrease with increasing ratios of NaReO$_4$ and NaVO$_3$ in the mixed salts. For the Re-doped MoS$_2$ monolayers, a drastic decrease of grain size is observed, from ~100 μm, ~20 μm to <10 μm with the $X_{ReMo}^{Re}$ increases from 5% to 10% and >25%, respectively. In contrast, the grain size of V-doped MoS$_2$ monolayers shows less sensitivity. Large grains of ~100 μm in size can be grown even at a high $X_{VMo}^{V}$ of 50%. **Figure 2**a shows a group photo of the Re and V-doped MoS$_2$ monolayers transferred onto double-side polished sapphire substrates. A gradual color change from light yellowish green (MoS$_2$ monolayers) to light brown color is noted when the $X_{ReMo}^{Re}$ changes from 0 to 50%. It matches well with the absorption spectra, where both A and B-exciton peaks are gradually decreasing and disappeared at 50% $X_{ReMo}^{Re}$ (Figure 2b). As a result, the photoemission peaks from A excitons are also decreased and red-shifted in the PL spectra (Figure 2c). This is because the substitutional Re doping changes the band structure of MoS$_2$ monolayers. New defect states are introduced near the conduction band minimum.[29, 39] Similarly, a gradual color change is observed in the V-doped MoS$_2$ monolayers (Figure 2a). The A and B-excitons absorption peaks have totally diminished in the V-doped MoS$_2$ monolayers that grown with 25% and higher $X_{VMo}^{V}$. More interestingly, a strong emission peak at lower energy of ~1.63 eV with a broad FWHM of ~ 200 meV (Figure 2c) is observed in the V-doped MoS$_2$ monolayers grown with 5% $X_{VMo}^{V}$. We attribute this new peak to the defect states above the valence band maximum caused by V doping into the lattice of MoS$_2$ monolayers.[40, 41] However, when the $X_{VMo}^{V}$ increases to 10%, only a very weak A-exciton emission and a broad low-energy defect-related emission are observed. The PL is fully quenched beyond 25% $X_{VMo}^{V}$, suggesting a heavily doped degenerate state is achieved.

Figure 2d demonstrates the evolution of Raman spectra for the Re and V-doped MoS$_2$ monolayers. The intrinsic MoS$_2$ monolayers have two characteristic Raman modes, namely E$_{2g}^1$ at ~386 cm$^{-1}$ and A$_{1g}$ at ~404 cm$^{-1}$ with a peak separation of ~18 cm$^{-1}$.[20] For the Re-doped

MoS$_2$ monolayers, new Raman modes recorded at 150, 211, 276 and 305 cm$^{-1}$ are ascribed to the ReS$_2$ monolayers.[43] These new Raman modes indicate the formation of ReS$_2$ domains in the Re-doped MoS$_2$ monolayers grown with $\geq 25\%$ $X_{ReMo}^{Re}$. Whereas for the V-doped MoS$_2$ monolayers, new Raman modes appeared at 158, 187, 227, 323 and 351 cm$^{-1}$ indicate the successful substitutional V doping in the MoS$_2$ lattice.[41]

Similarly, we also conducted the CVD growth of Re and V-doped WS$_2$ and WSe$_2$ monolayers. Their optical images are shown in Figure S3. Their Raman spectra show solid evidence for the successful growth of Re and V-doped WS$_2$ and WSe$_2$ monolayers. New Raman modes and quenched PL are observed in these samples (Figure S4). However, it is hard to grow Re-doped WS$_2$ and WSe$_2$ monolayers due to the large formation energy.[44] In the future, more efforts should be given to the growth of n-type Re-doped TMDC monolayers.[45]

(**Figure 3**)

The atomic structures of Re and V-doped TMDC monolayers were investigated by atomic-resolution scanning transmission electron microscope (STEM). **Figure 3**a-d show four typical annular dark field (ADF) STEM images of Re and V-doped TMDC monolayers: Re-doped MoS$_2$, V-doped MoS$_2$, V-doped WS$_2$ and V-doped WSe$_2$, respectively. As shown in Figure 3a, Re atoms show brighter contrast than the Mo atoms since the ADF contrast is propotional to Z$^2$ (Re (Z=75), Mo (42), S (16)). In most area of the Re-doped MoS$_2$ monolayers (grown with 25% $X_{ReMo}^{Re}$), the Re dopants are uniformly dispersed in the MoS$_2$ lattice with a concentration of ~2.1%, which is probably limited by the solubility. On the other hand, we have also found small ReS$_2$ domains containing Mo dopants, in a size of ~150 $\times$ 150 nm$^2$, embedded in the host MoS$_2$ (Figure S5). Note that the ReS$_2$ presents distorted 1T phase. Such phase separation is more pronounced at higher Re doping concentration when the $X_{ReMo}^{Re}$ exceeds 25%. This matches well with the observation of enhanced ReS$_2$ Raman modes in the Re-doped MoS$_2$ monolayers grown with 25% and 50% $X_{ReMo}^{Re}$ (Figure 2d).

Figure 3b-d show the ADF-STEM images of V-doped MoS$_2$, WS$_2$ and WSe$_2$ monolayers, respectively. The V dopants are uniformly dispersed in these samples without phase segregation, indicating a good miscibility of V atoms in these TMDC monolayers.[34, 35, 46] From the STEM images, we estimate ~2.9%, ~4.7% and ~ 2.7% of V atoms are doped in 25% $X_{VMo}^{V}$-MoS$_2$, 5% $X_{VW}^{V}$-WS$_2$ and 5% $X_{VW}^{V}$-WSe$_2$ monolayers, respectively. Compared to the W-based TMDCs, the doping concentration of V in MoS$_2$ monolayers is not as high (or merely equivalent) even with an increased amount of NaVO$_3$ (25%) used. The result implies a higher compability of V in WS$_2$ than WSe$_2$ over MoS$_2$.

(**Figure 4**)

Substitutional replacement of Mo or W with electron donor (Re) and acceptor (V) can dramatically change the electronic structure, carrier types and conductivity of TMDC monolayers. **Figure 4a** demonstrates the typical FET transport curves of Re and V-doped MoS$_2$ monolayers. With increasing NaReO$_4$ ratio in the mixed salts, we see a dramatic shift of threshold voltage to more negative gate bias, indicating a strong electron doping in the as-grown Re-doped MoS$_2$ monolayers. Compared to the intrinsic MoS$_2$ monolayers, the on-state current and current on/off ratio decreased by two orders of magnitude (from ~10$^{-5}$ A to 10$^{-7}$ A; 10$^8$ to 10$^6$) for 5% $X_{ReMo}^{Re}$-MoS$_2$ monolayers, which mainly due to the scattering caused by the Re atoms in MoS$_2$ lattice. Degenerate electron transport and improved conductivity are achieved in 10% and 25% $X_{ReMo}^{Re}$-MoS$_2$ monolayers. For the V-doped MoS$_2$ monolayers, the on-state current for electron transport shows a steady decrease but still maintain a high current on/off ratio up to ~10$^7$ for 5% and 10% $X_{VMo}^{V}$-MoS$_2$ monolayers. Meanwhile, the threshold voltage shows a positive shift due to the strong hole doping. A degenerate doping is achieved in 25-75% $X_{VMo}^{V}$-MoS$_2$ monolayers, poor gate-tunability and metallic transport behavior are observed in these samples.

Figure 4b shows the typical FET transport curves of V-doped WS$_2$ and WSe$_2$ monolayers. Intrinsic WS$_2$ and WSe$_2$ monolayers show poor transport properties when Au was used as contacts. With increasing ratios of NaVO$_3$, a steady increment of hole conductivity is achieved in the V-doped samples. The conductivity in 10-25% $X_{VW}^{V}$-WS$_2$ (WSe$_2$) monolayers show a dramatic increase up to $10^8$ times, compared to the intrinsic WS$_2$ and WSe$_2$ at zero gate bias. This matches well with the results of density functional theory (DFT) calculations. With the increase of the doping concentration of V atoms in WSe$_2$ monolayers, a steady down-shift of the Fermi level into the valence band is observed (Figure S6), indicating the increase of hole concentration. The calculated partial charge density associated with states near the Fermi level in 11.1% V-doped WSe$_2$ (Figure S6e) are quite delocalized compared to that in 1% V-doped WSe$_2$ (Figure S6c), which localized around the dopants. This indicates a higher hole conductivity in the 11.1% V-doped WSe$_2$. In addition, much higher hole conductivity is achieved in the V-doped WS$_2$ and WSe$_2$ than V-doped MoS$_2$. We attribute this to the feasible high concentration of ionized V dopants in WSe$_2$ and WS$_2$ monolayers (Figure 3c-d and Figure S6). Meanwhile, high ratio of NaVO$_3$ (25% $X_{VW}^{V}$) leads to the growth of multilayer V-doped WS$_2$ and highly reactive V-doped WSe$_2$ monolayers. The 25% $X_{VW}^{V}$-WSe$_2$ monolayers generate holes soon after exposed to the ambient atmosphere, which accounts for the similar electrical properties observed in the 25% and 10% $X_{VW}^{V}$-WSe$_2$ monolayers.

For the contact engineering of 2D electronics, the metallic 10% $X_{VW}^{V}$-WSe$_2$ monolayers was employed as vdW contacts for WSe$_2$ FETs due to their high conductivity. Figure 4c-d illustrate two types of contacts for WSe$_2$-FETs: one is the widely used Au or Pd metal contact (Figure 4c). The other is the degenerately V-doped WSe$_2$ contact. An optical image showing the vdW contacts formed by transferring the etched V-doped WSe$_2$ monolayers onto WSe$_2$ monolayers is depicted in Figure 4d. Detailed fabrication process is presented in Figure S7 and experimental section. Figure 4e is the typical transport curves of WSe$_2$-FETs with three different contacts,

Au, Pd and V-doped WSe$_2$. When Au was employed as contact, the WSe$_2$-FETs show an obvious ambipolar transport behavior with on-state current of ~10 nA and a current on/off ratio of ~10$^4$. Pd is ideal for hole transport in WSe$_2$-FETs due to its high work function. Changing the contact metal from Au to Pd has improved the on-state current for hole transport by two orders of magnitude (from ~10$^{-8}$ A to 10$^{-6}$ A). Meanwhile, when V-doped WSe$_2$ was used as contacts, the on-state current for hole transport is enhanced by three orders of magnitude compared to Au (from ~10$^{-8}$ A to 10$^{-5}$ A), also showing a much better result than the Pd contact. This enables an improved current on/off ratio of V-doped WSe$_2$ contacted WSe$_2$-FETs, reaching a high value of 10$^8$. Furthermore, compared to the metal contacts, the electron transport can be fully quenched at positive gate bias in V-doped WSe$_2$ contact. This implies a promising application in low-power consumption nanoelectronics.

To better understand the difference between metal (Au) and V-doped WSe$_2$ vdW contact for WSe$_2$-FETs, corresponding contact geometries were built and simulated using the first-principles DFT calculations (Figure S8). For the Au-WSe$_2$ contact, the calculated projected density of states (PDOS) indicates that Au can metallize WSe$_2$ monolayer strongly and fill the bandgap with states (Figure 4f). This further confirms that the metal-induced gap states (MIGS) are formed in the contact between traditional metals and TMDC monolayers. The MIGS and large strain at the interface result in the large contact resistance.[6-12] Therefore, instead of the conventional metal deposition process, Au eletrodes were transferred as 3D vdW contacts in the TMDC-based electronic devices to elimate the MIGS and large strains generated.[25] In stark contrast to the Au contact, the the WSe$_2$-FETs show p-type behaviour with the V-doped WSe$_2$ vdW contact. The smaller metallization effect preserves the intrinsic bandgap of WSe$_2$ monolayer (Figure 4g and Figure S6b). Our simulation results well explained the experimental observations, suggesting the V-doped WSe$_2$ monolayers as promising electrode material for the p-type WSe$_2$-FETs. In addition, because of the similar lattice parameters, the negligible distortion/strain at junction of the intrinisic and V-doped WSe$_2$ monolayers has contributed to the low-resistance contacts.[12]

In summary, CVD method using mixed molten salts is highly promising for the VLS growth of Re and V-doped TMDC monolayers. Tunable composition, optical and electrical properties are achieved in the Re and V-doped TMDC monolayers. The metallic V-doped WSe$_2$ monolayers are ideal p-type vdW contact for the WSe$_2$-FETs. Much improved device performance is observed compared to the traditional Au and Pd contacts. Meanwhile, in order to obtain a good contact for electron transport, more efforts should be devoted to the growth of degenerate Re-doped TMDCs or explore new low-work-function 2D metals. These Re and V-doped TMDCs are expected to bring profound impacts to the 2D TMDC-based electronics as promising electrodes and interconnects. In addition, as 2D lateral contacts and PN junctions can also be created using patterned growth with molten slats, this will further improve the incorporation and functionality of 2D TMDC-based integrated circuits.

**Experimental Section**

*Preparation of mixed salt precursors*: First, 20 mM Na$_2$MoO$_4$ (99%, Sterm Chemicals), Na$_2$WO$_4$ (99+%, Strem Chemicals), NaReO$_4$ (99.95%, Alfa Aesar) and NaVO$_3$ (96%, Alfa Aesar) aqueous solutions were prepared by dissolving the salts in DI water, respectively. Then, the aqueous salt solutions were mixed in designated ratios for growing Re and V-doped TMDC monolayers. One-side polished sapphire substrates (c-plane, Shinkosha) were treated with UV-O$_3$ for 30 min to obtain hydrophilic surface. The mixed salt solutions were then spin-coated on the treated sapphire substrates with a speed of 5000 rpm for 30 seconds.

*CVD growth of Re and V-doped TMDC monolayers*: All the CVD growths were conducted in a 2-inch tube furnace. The temperature ramping rate was 30 ºC/min. Crucibles containing sulphur (99.999%, Fujifilm) and selenium (99.99%, Strem Chemicals) were kept at ~180 ºC and ~300 ºC during growth, respectively. To avoid possible contamination, each quartz tube is assigned for growing one specific Re and V-doped TMDCs. For Re and V-doped $MoS_2$ monolayers, the growth was performed at 750 °C for 10 minutes with 200 sccm high-purity argon as carrier gas. For Re and V-doped $WS_2$ and $WSe_2$ monolayers, the growth was performed at 800 °C for 10 minutes with 200 sccm high-purity $Ar/H_2$ (5%) forming gas.

*Raman and PL*: First, the Re and V-doped TMDCs grown on sapphire substrates were transferred onto $SiO_2$ (285 nm)/Si substrates. Then, Raman and PL measurements were performed using a laser confocal microscope (Tokyo Instruments, Nanofinder FLEX). A 532 nm excitation laser with a spot size of 1 μm was focused onto the sample surface. The samples' Raman/PL signals were detected by an electron multiplying CCD detector through a grating with 2400 grooves/mm for Raman and 150 grooves/mm for PL, respectively.

*STEM and EELS*: STEM images were acquired by using JEOL 2100F microscope equipped with a JEOL-DELTA correctors and the cold field emission gun operating at 60 kV. The probe current was about 25-30 pA. The convergence semiangle was 35 mrad and the inner acquisition semiangle was 79 mrad. The EELS core loss spectra were taken by using Gatan low-voltage quantum spectrometer.

*FET fabrication and measurements*: The Re and V-doped TMDC monolayers were transferred onto $SiO_2$ (285 nm)/Si substrates first. Then, a LED photolithography and oxygen plasma etching were conducted on Re and V-doped TMDC monolayers to define the channel shape. Next, another LED photolithography was repeated to pattern the electrodes. E-beam evaporator was utilized to deposit Cr/Au (1/50 nm) as contact. A standard lift-off process was employed by rinsing the substrates in acetone and IPA sequentially. All the transistors have the same channel size with length: ~20 μm and width: ~4 μm. Measurements were carried out in a high vacuum of ~$2\times10^{-4}$ Pa. The backgate bias ($V_{gs}$) was swept between -30 V and 30 V with a step of 1 V. The source-drain bias ($V_{ds}$) is 1 V.

*Fabrication of V-doped $WSe_2$ contacted $WSe_2$-FETs*: First, parallel gaps were fabricated on the 10% $X_{VW}^V$-$WSe_2$ monolayers using LED photolithography (patterning) and etched with oxygen plasma sequentially. Then, the etched V-doped WSe$_2$ monolayers were transferred onto CVD-gown WSe$_2$ monolayers. Next, the second patterning and etching process was performed to define the channel area. Finally, photolithography was applied to pattern the electrodes and Au (50 nm) film was deposited using e-beam evaporator (Figure S7).

*DFT simulation:* The DFT simulation was performed within the generalized gradient approximation of the Perdew, Burke, and Ernzernhof functional as implemented in the Vienna ab initio simulation package.[47, 48] Electron-ion interactions were described using projector-augmented wave pseudopotentials.[49] The effective Hubbard U value was set to 4.2 eV for the V 3d state.[50-52] In the calculation of V-doped WSe$_2$/WSe$_2$ and Au/WSe$_2$ contacts, the Grimme's DFT-D2 method is employed for vdW correction.[53] A plane-wave kinetic-energy cutoff of 600 eV and a k-spacing of 0.2 Å$^{-1}$ in reciprocal space were used to ensure that the energy converged to better than 1 meV/atom.

**Supporting Information**
Supporting Information is available from the Wiley Online Library or from the author.

**Acknowledgements**
S.L. acknowledges the support from JSPS-KAKENHI (19K15399). T.T. acknowledges the support from JSPS-KAKENHI (17K19187). Y.-C. Lin and K.S. acknowledge support from the JSPS-KAKENHI (16H06333), (18K14119), JSPS A3 Foresight Program, and Kazato Research Encouragement Prize. B.G. and Y.T. thank the support by MEXT as "Program for Promoting Researches on the Supercomputer Fugaku (Fugaku Battery & Fuel Cell Project), Grant JPMXP1020200301, and the supercomputer at NIMS. H.E.L. acknowledges the support from JSPS-KAKENHI (19K15393). S.L. acknowledges all staff members of the Nanofabrication group at NIMS for their support.

Received: ((will be filled in by the editorial staff))
Revised: ((will be filled in by the editorial staff))
Published online: ((will be filled in by the editorial staff))

References

[1] Q. H. Wang, K. Kalantar-Zadeh, A. Kis, J. N. Coleman, M. S. Strano, *Nat. Nanotechnol.* **2012**, 7, 699.

[2] B. Liu, A. Abbas, C. Zhou, *Adv. Electron. Mater.* **2017**, 3, 1700045.

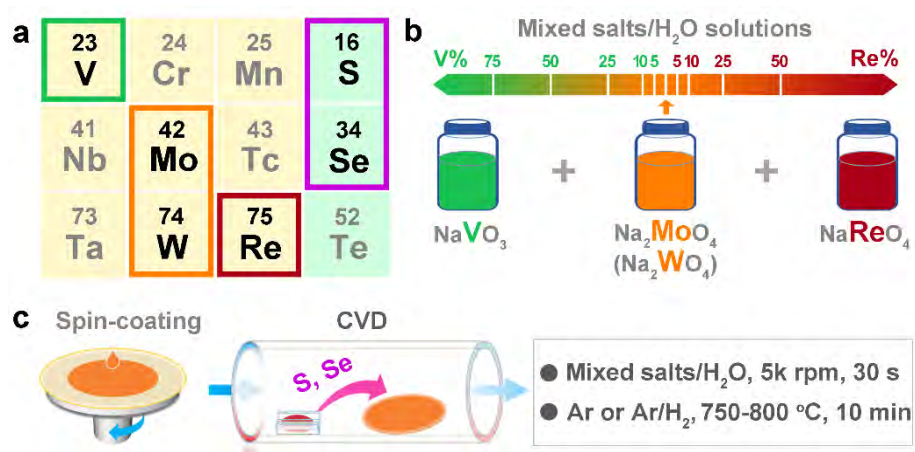

**Figure 1.** Strategy for the CVD growth of 2D Re and V-doped TMDCs. (a) A partial periodic table indicating the substitutional Re (electron donor) and V (electron acceptor) doping of 2D TMDCs ($MS_2$, and $MSe_2$, $M$ = Mo, W). (b) Mixed salt solutions with different $NaReO_4$ and $NaVO_3$ ratios are prepared as source precursors. (c) Schematic illustrations of the spin-coating of mixed salt solution onto growth substrate and the conditions employed in the CVD growth of Re and V-doped TMDC monolayers.

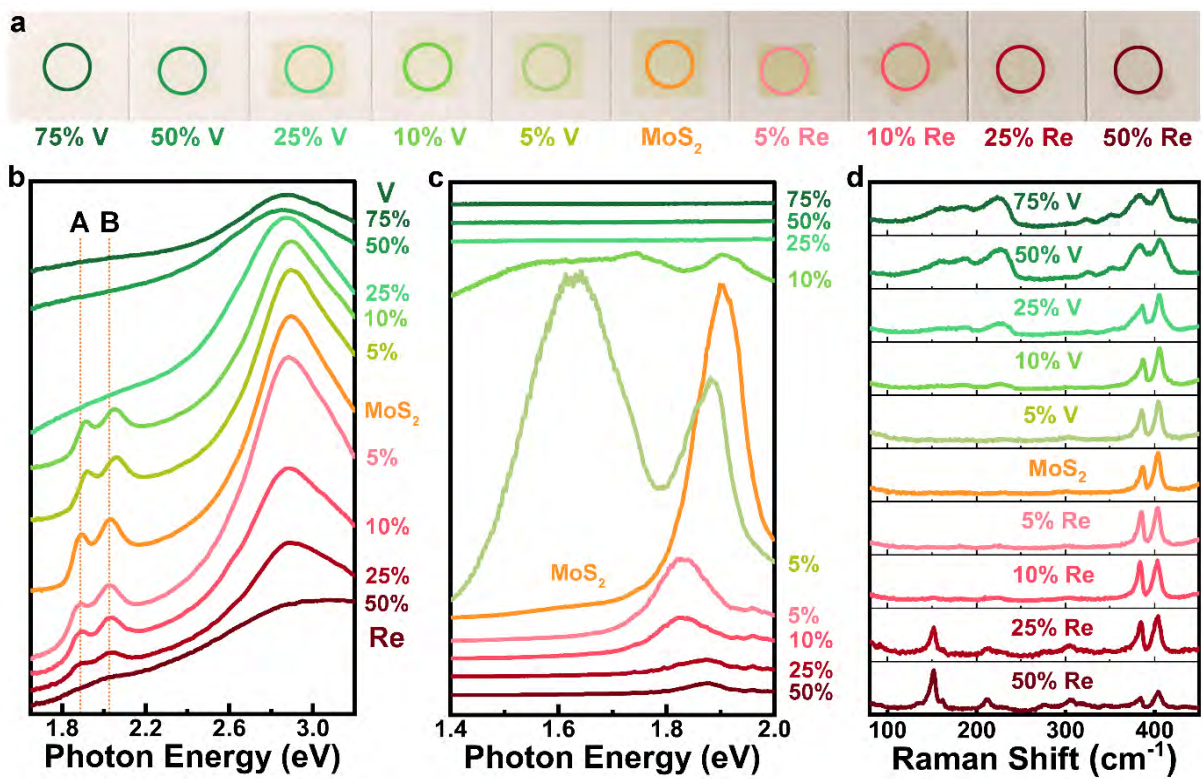

**Figure 2.** Spectroscopic characterization of Re and V-doped MoS$_2$ monolayers. (a) Optical images of the transferred Re and V-doped MoS$_2$ monolayers on double-side polished sapphire substrates. The circles indicate the area inspected in UV-Vis spectroscopy. (b) Absorption spectra of the transferred Re and V-doped MoS$_2$ monolayers shown in (a). (c, d) Typical (c) photoluminescence and (d) Raman spectra of the Re and V-doped MoS$_2$ monolayers transferred on SiO$_2$/Si substrates. All the Re% and V% labelled in (a-d) represent the NaReO$_4$ and NaVO$_3$ ratios in the mixed salt solutions, not the actual Re and V concentrations in the doped MoS$_2$ monolayers. In figures b-d, y axis of the plots represent intensity with arbitrary unit.

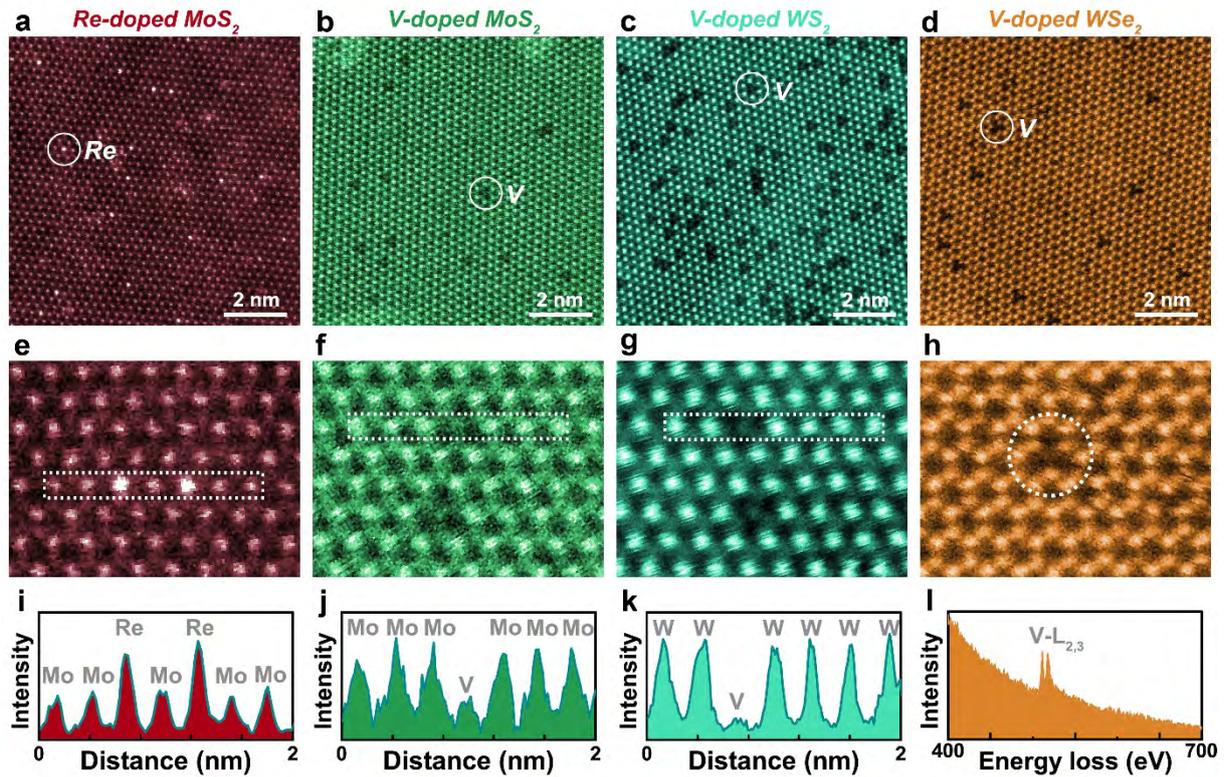

**Figure 3.** Atomic structures of Re and V-doped TMDC monolayers. (a-d) Low magnification ADF-STEM images of (a) Re-doped MoS$_2$ (25% $X^{Re}_{ReMo}$), (b) V-doped MoS$_2$ (25% $X^{V}_{VMo}$), (c) V-doped WS$_2$ (5% $X^{V}_{VW}$) and (d) V-doped WS$_2$ (5% $X^{V}_{VW}$), respectively. (e-h) The corresponding high magnification ADF-STEM images. (i-k) The ADF intensity profiles of Re dopants in MoS$_2$, V dopants in MoS$_2$ and WS$_2$ extracted from the dotted boxes in (e-g) identifying the position of Re and V from the intensity. (l) The EELS spectrum of V dopant in WSe$_2$ taken from the dotted circle in (h).

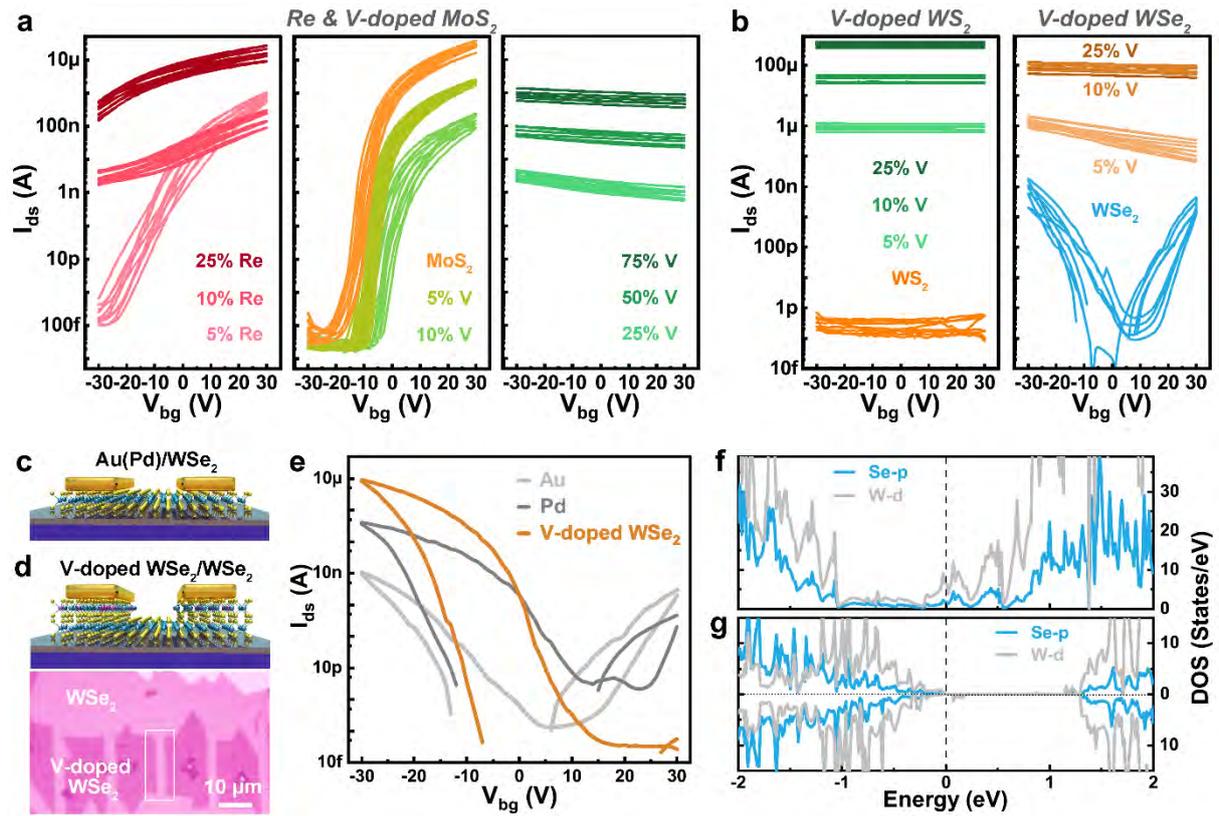

**Figure 4.** Tunable electrical properties of Re and V-doped TMDCs and contact engineering of WSe$_2$-FETs. (a) Typical transport curves of Re and V-doped MoS$_2$ monolayers. (b) Typical transport curves of V-doped WS$_2$ and WSe$_2$ monolayers. All the Re% and V% represent the NaReO$_4$ and NaVO$_3$ ratios in the mixed salt solutions. (c) Schematic of Au or Pd contact for WSe$_2$-FETs. (d) Schematic and optical image of V-doped WSe$_2$ as vdW contact for a WSe$_2$-FET. (e) Comparison of transport properties of WSe$_2$-FETs with three kinds of contacts: Au, Pd and V-doped WSe$_2$. (f, g) PDOS of WSe$_2$ monolayers contacting with (f) Au and (g) V-doped WSe$_2$, respectively.

Supporting Information

**Tunable Doping of Rhenium and Vanadium into Transition Metal Dichalcogenides for Two-Dimensional Electronics**

*Shisheng Li\*, Jinhua Hong, Bo Gao, Yung-Chang Lin, Hong En Lim, Xueyi Lu, Jing Wu, Song Liu, Yoshitaka, Tateyama, Yoshiki Sakuma, Kazuhito Tsukagoshi, Kazu Suenaga and Takaaki Taniguchi\**

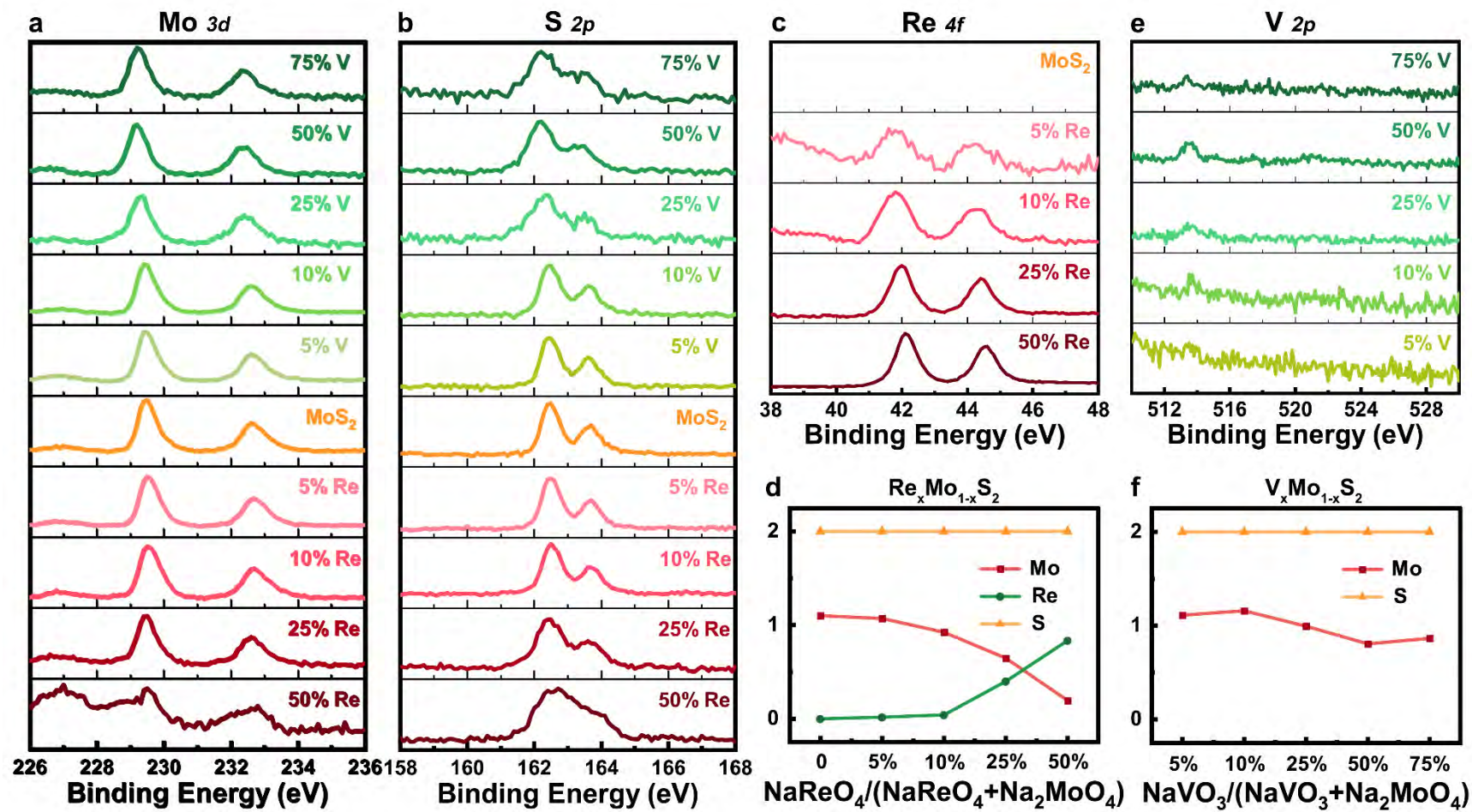

**Figure S1. XPS spectra of Re and V-doped MoS$_2$ monolayers.** XPS fine spectra of (a) Mo 3d, (b) S 2p, (c) Re 4f and (e) V 2p. (d) The evolution of Re and Mo atomic ratios in the as-grown Re-doped MoS$_2$ monolayers. (f) The evolution of Mo atomic ratios in the as-grown V-doped MoS$_2$ monolayers. S is calibrated to 2 in (d) and (f) for comparison. In figures a-c and e, y axis of the plots represent intensity with arbitrary unit.

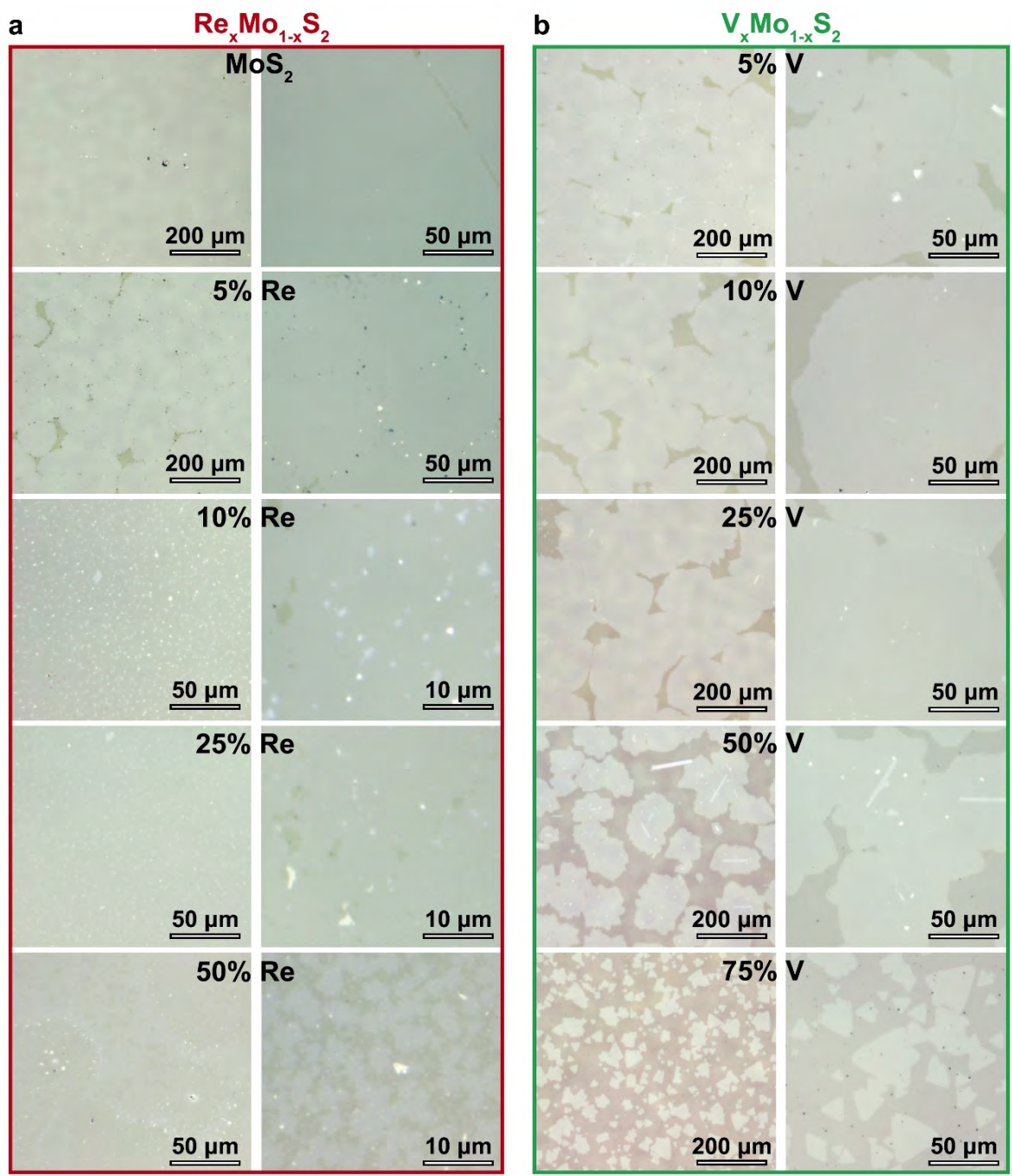

**Figure S2. Optical images of Re and V-doped MoS₂ monolayers.** (a) The ratios of NaReO₄ in mixed salt precursors, $X_{ReMo}^{Re}$ are 0, 5%, 10%, 25% and 50%. (b) The ratios of NaVO₃ in mixed salt precursors, $X_{VMo}^{V}$ are 5%, 10%, 25%, 50% and 75%.

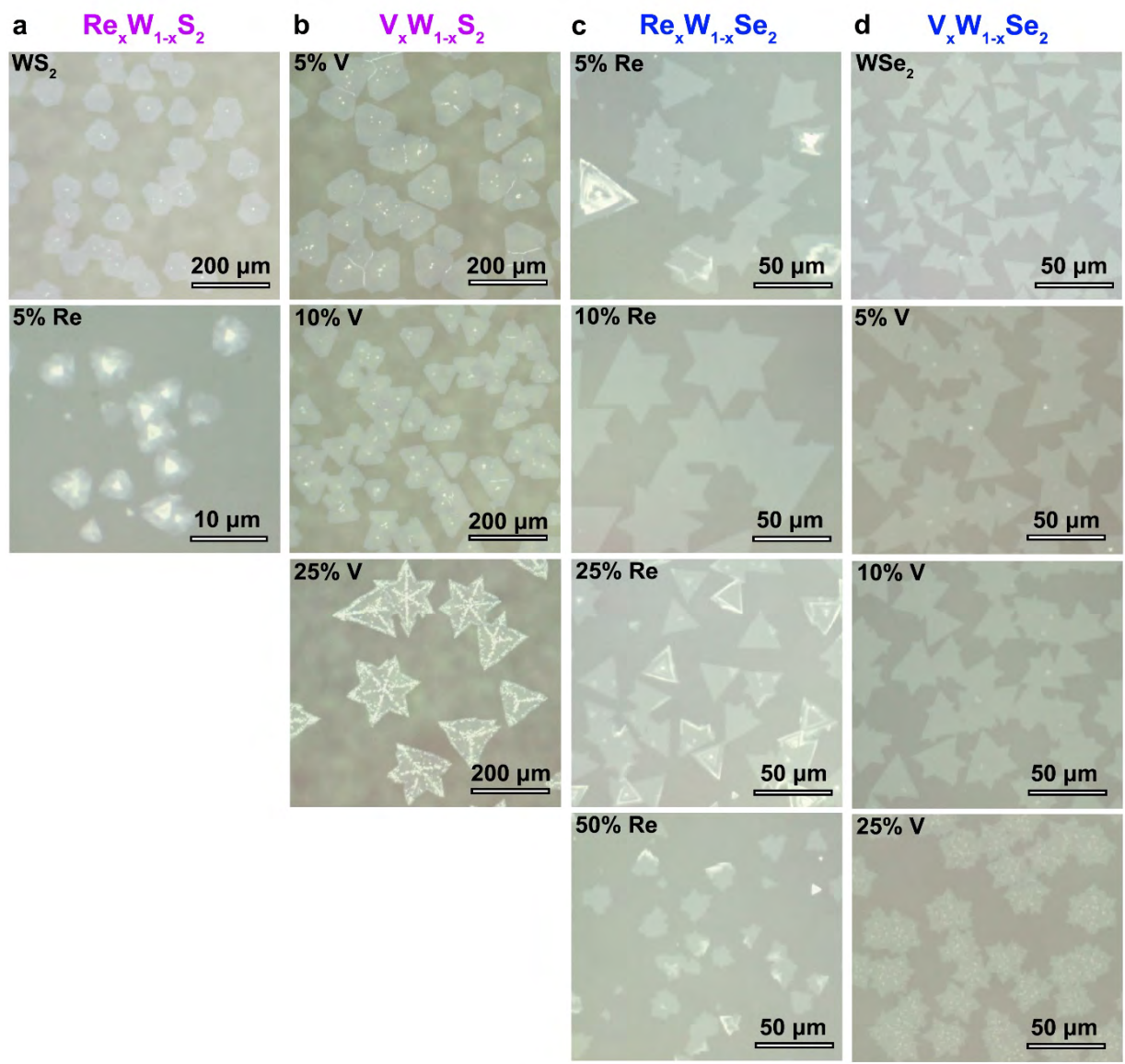

**Figure S3. Optical images of Re and V-doped WS$_2$ and WSe$_2$.** The ratios of dopant salts in mixed salt precursors: (a) $X^{Re}_{ReW}$ are 0 and 5%. (b) $X^{V}_{VW}$ are 5%, 10% and 25%; (c) $X^{Re}_{ReW}$ are 5%, 10%, 25% and 50%; (d) $X^{V}_{VW}$ are 0, 5%, 10% and 25%.

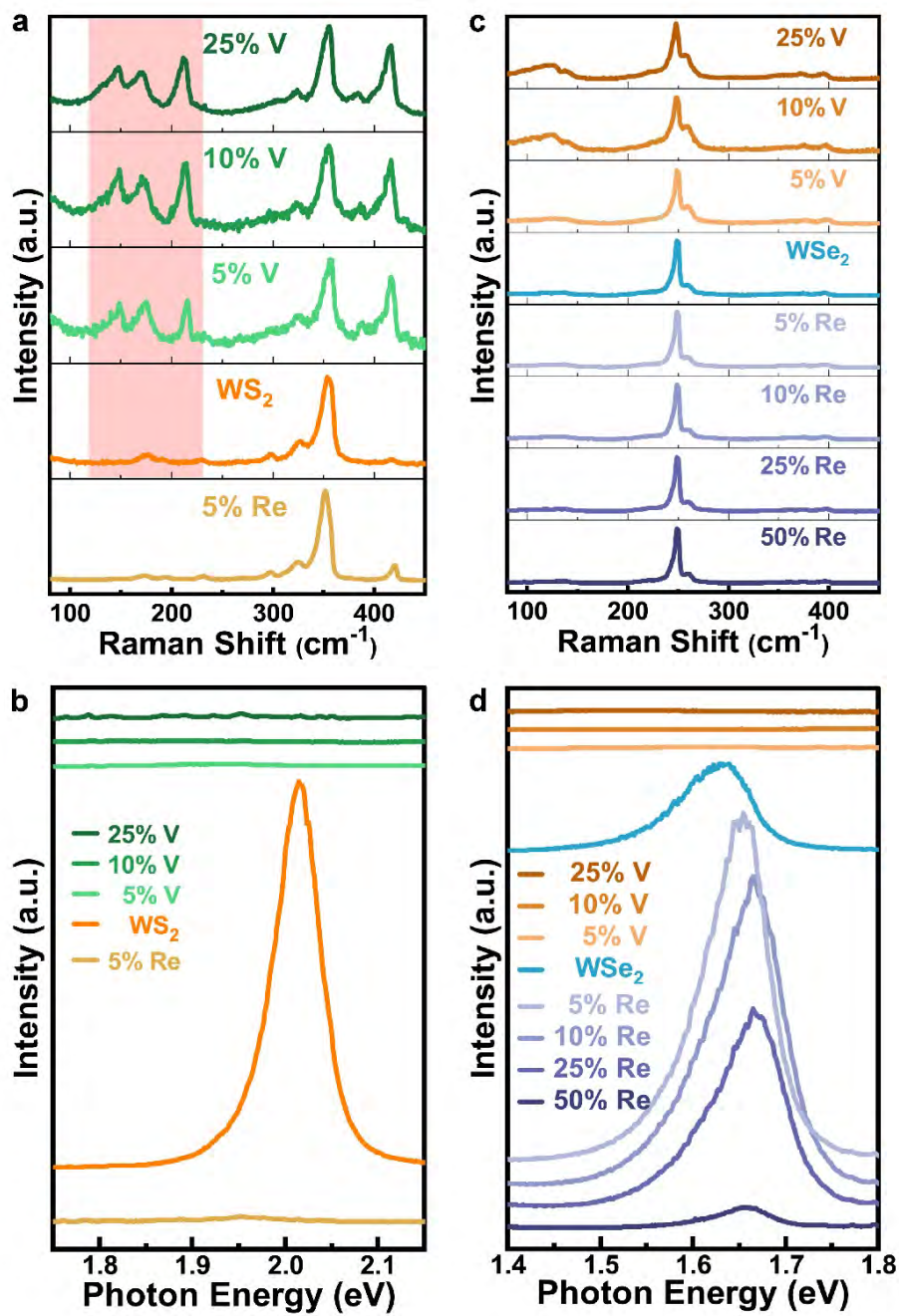

**Figure S4. Raman and PL spectra of Re and V-doped WS$_2$ and WSe$_2$ monolayers.** (a) Raman and (b) PL spectra of Re and V-doped WS$_2$ monolayers. (The 5% $X_{ReW}^{Re}$-WS$_2$ are multilayers). (c) Raman and (d) PL spectra of Re and V-doped WSe$_2$ monolayers.

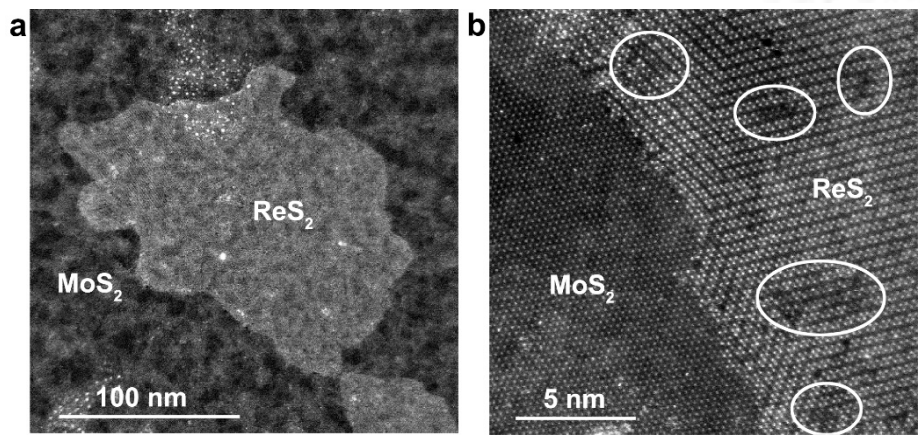

**Figure S5. Structural characterization of 25% $X_{ReMo}^{Re}$ -MoS₂ monolayers.** (a) Low-magnification STEM image of a ReS₂ domain in MoS₂ matrix. (b) High-magnification STEM image showing the atomic structures of Re-doped MoS₂ and ReS₂. White circles indicate the substitutional Mo doping in the ReS₂ domain.

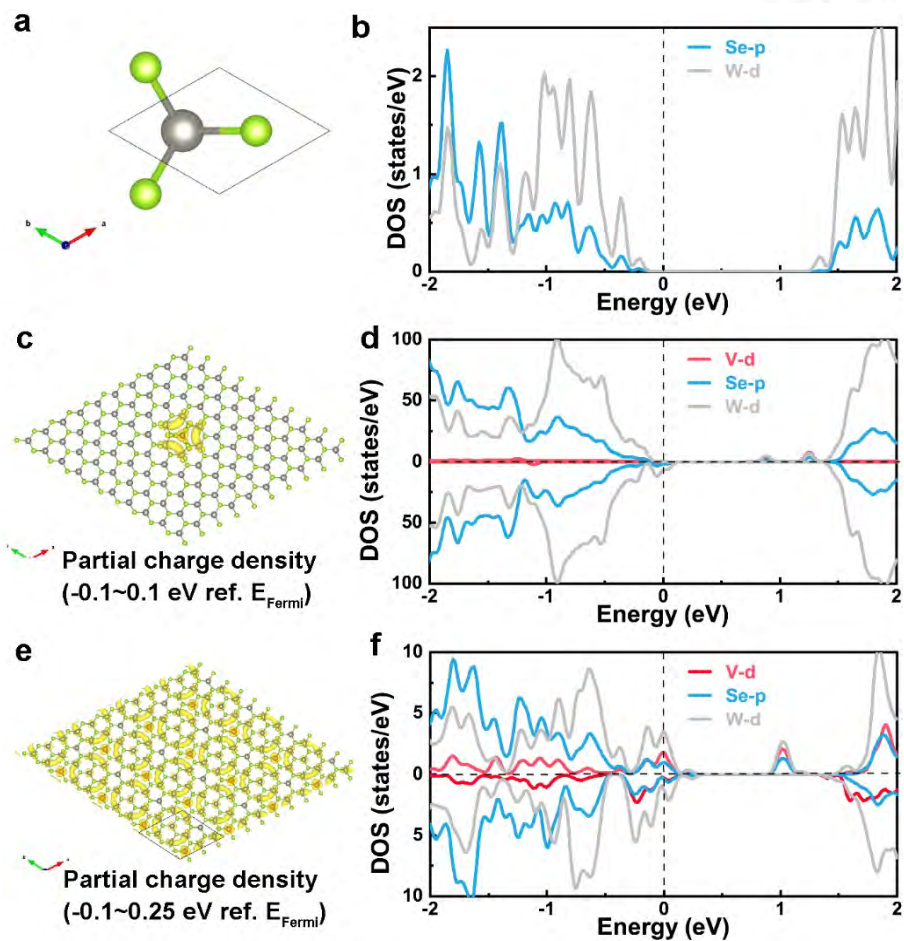

**Figure S6. Charge densities and PDOSs of V-doped WSe$_2$ monolayers.** (a) Unit cell and (b) PDOS of WSe$_2$ monolayer. (c) Partial charge density associated with the states from -0.1 to 0.1 eV relative to the Fermi level and (d) PDOS of 1 % V-doped WSe$_2$ monolayer. (e) Partial charge density associated with the states from -0.1 to 0.25 eV relative to the Fermi level and (f) PDOS of 11.1 % V-doped WSe$_2$ monolayer. In the PDOSs figures, the dashed lines indicate the Fermi level.

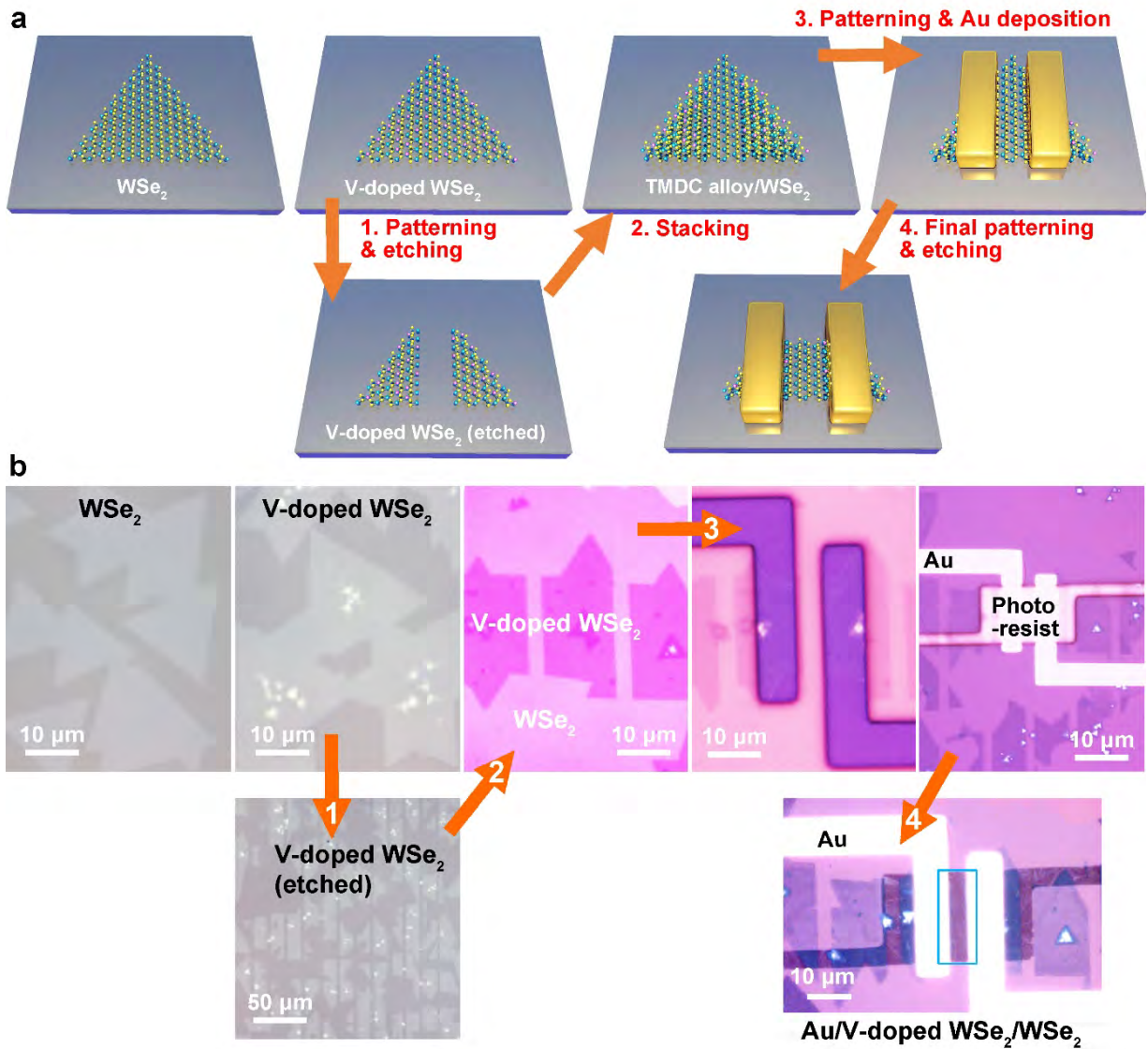

**Figure S7. Fabrication of WSe$_2$-FETs with V-doped WSe$_2$ vdW contact.** (a) Schematics and (b) corresponding optical images of the detailed fabrication process for 10% $X_{VW}^{V}$-WSe$_2$ monolayer contacted WSe$_2$-FETs.

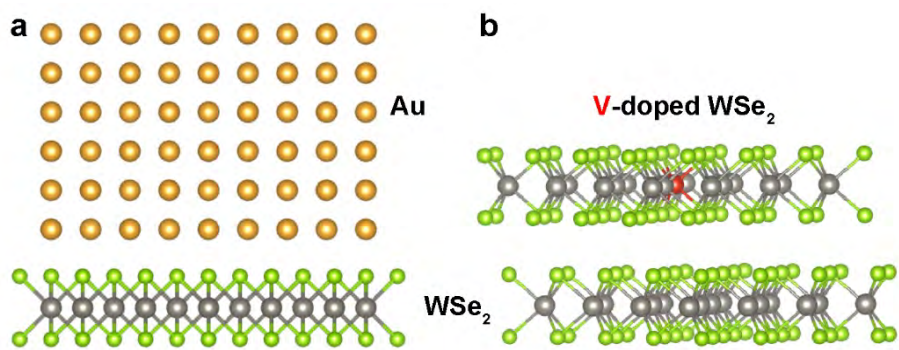

**Figure S8. Two contact geometries in the DFT simulation.** (a) Au/WSe$_2$ contact. (b) V-doped WSe$_2$/WSe$_2$ contact.